\documentclass[12pt]{iopart}
\usepackage{graphicx,amssymb}
\begin{document}

\title[Vector meson production with the ALICE detector]{Vector meson production in pp collisions at $\sqrt{s}=7\rm{~TeV}$, measured with the ALICE detector}

\author{A. De Falco for the ALICE collaboration}

\address{Universit\`a di Cagliari and INFN Sezione di Cagliari, Italy}
\ead{alessandro.de.falco@ca.infn.it}
\begin{abstract}
\label{abstract}
Vector mesons are key probes of the hot and dense state of strongly interacting matter 
produced in heavy ion collisions. Their dileptonic decay channel is particularly suitable 
for these studies, since dileptons have negligible final state interactions in hadronic matter. 
A preliminary measurement of the $\phi$ and $\omega$ differential cross sections 
was performed by the ALICE experiment in pp collisions at $\sqrt{s}=7$~TeV, through their decay in muon pairs. 
The $p_{\rm T}$ and rapidity regions covered in this analysis are $p_{\rm T}>1$~GeV$/c$ and $2.5 < y < 4$. 
\end{abstract}
Low mass vector meson ($\rho^0, \omega, \phi$) production provides key 
information on the hot
and dense state of strongly interacting matter produced 
in high-energy heavy ion collisions. Among them,
strangeness enhancement can be accessed through the measurement
of $\phi$ meson production, while the measurement of the $\rho$ 
spectral function can be used to reveal in-medium modifications of 
hadron properties close to the QCD phase boundary. 
Vector meson production in pp collisions provides a reference 
for these studies. 
Moreover, it is interesting by itself, since it can be used to tune
particle production models in the unexplored LHC energy
range.

The ALICE experiment at the LHC can access 
vector mesons produced at forward rapidity through their decays 
in muon pairs, and at central rapidity in the di-electron decay channel.
The detector is fully described in~\cite{ALICE-detector}.
In this paper, results from the analysis of the data collected during the 
2010 pp run at $\sqrt{s}=7$~TeV are reported. 

The measurement in the dimuon channel was performed using the 
forward muon spectrometer, 
that consists of an absorber acting as muon filter, a set of cathod pad
chambers (five stations, each one composed of two chambers) for the 
track reconstruction in a dipole field, two stations of two resistive
plate chambers for the muon trigger, two absorbers and an iron wall
acting as a muon filter.

The data sample used for the analysis in the dimuon channel amounts to 
an integrated luminosity of approximately 85~nb$^{-1}$. Since only a 
fraction of the data contained the full information relevant 
for the extraction of the integrated luminosity, 
a subsample corresponding to $L_{\mathrm{INT}}=55.7~\mathrm{nb}^{-1}$ 
was used for the measurement of the $\phi$ cross section, while the full 
sample was used to extract the $p_T$ distribution. Muon pairs were 
selected asking that each muon track reconstructed in the tracking chambers 
matches the corresponding tracklet in the trigger stations
in the position in the (x-y) plane and in the slope in the 
(r-z) plane. A cut on the muon 
rapidity $2.5<y_{\mu} < 4$ was applied in order to remove the tracks 
close to the acceptance borders. About 291,000 opposite sign $(N_{+-})$, 
197,000 like-sign $(N_{++},~N_{--})$ muon pairs survived these selections. 

The combinatorial background was evaluated using the event mixing technique, 
and normalized to $2 R \sqrt{N_{++}N_{--}}$, 
where $R=A_{+-}/\sqrt{A_{++}A_{--}}$, and $A_{\pm \pm}$ is the acceptance for 
a (${\pm \pm}$) pair. 
The event mixing was checked by comparing the results obtained for 
like-sign mixed pairs with the real ones. The shapes of the 
background calculated with the two methods are identical, 
while the amount of like-sign pairs estimated with the 
event mixing differs from the one in the real data by $5\%$. We take 
this value as the uncertainty on the background normalization.  
The signal-to-background ratio for $p_{\rm T}>1$~GeV/$c$ is 
about 1 at the $\phi$ and $\omega$ masses.
Alternatively, the combinatorial background contribution to the 
opposite sign mass spectrum for a given $\Delta M$ mass bin can be 
evaluated from the like sign mass spectra using the formula:
$N_{+-}^{\rm comb}(\Delta M) = 2 R \sqrt{N_{++}(\Delta M)N_{--}(\Delta M)}.$
The two techniques are in good agreement for $p_{\rm T} > 1$~GeV/$c$, 
while for lower pair transverse momenta 
both methods fail in describing the background. 
The analysis is thus limited to $p_{\rm T}>1$~GeV$/c$. 

\begin{figure}
	\centering
	\includegraphics[width=0.49 \columnwidth]{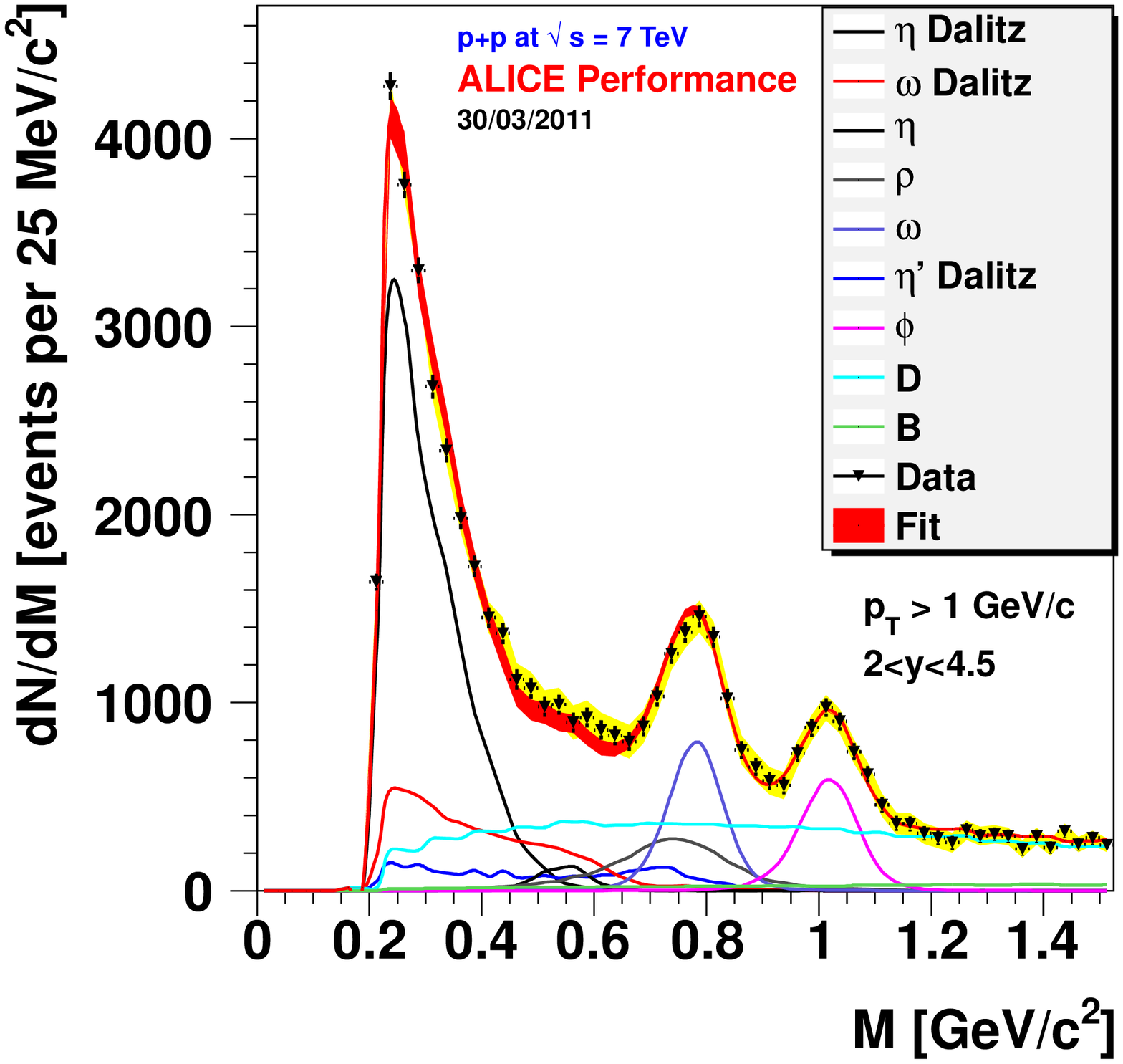}
	\includegraphics[width=0.49 \columnwidth]{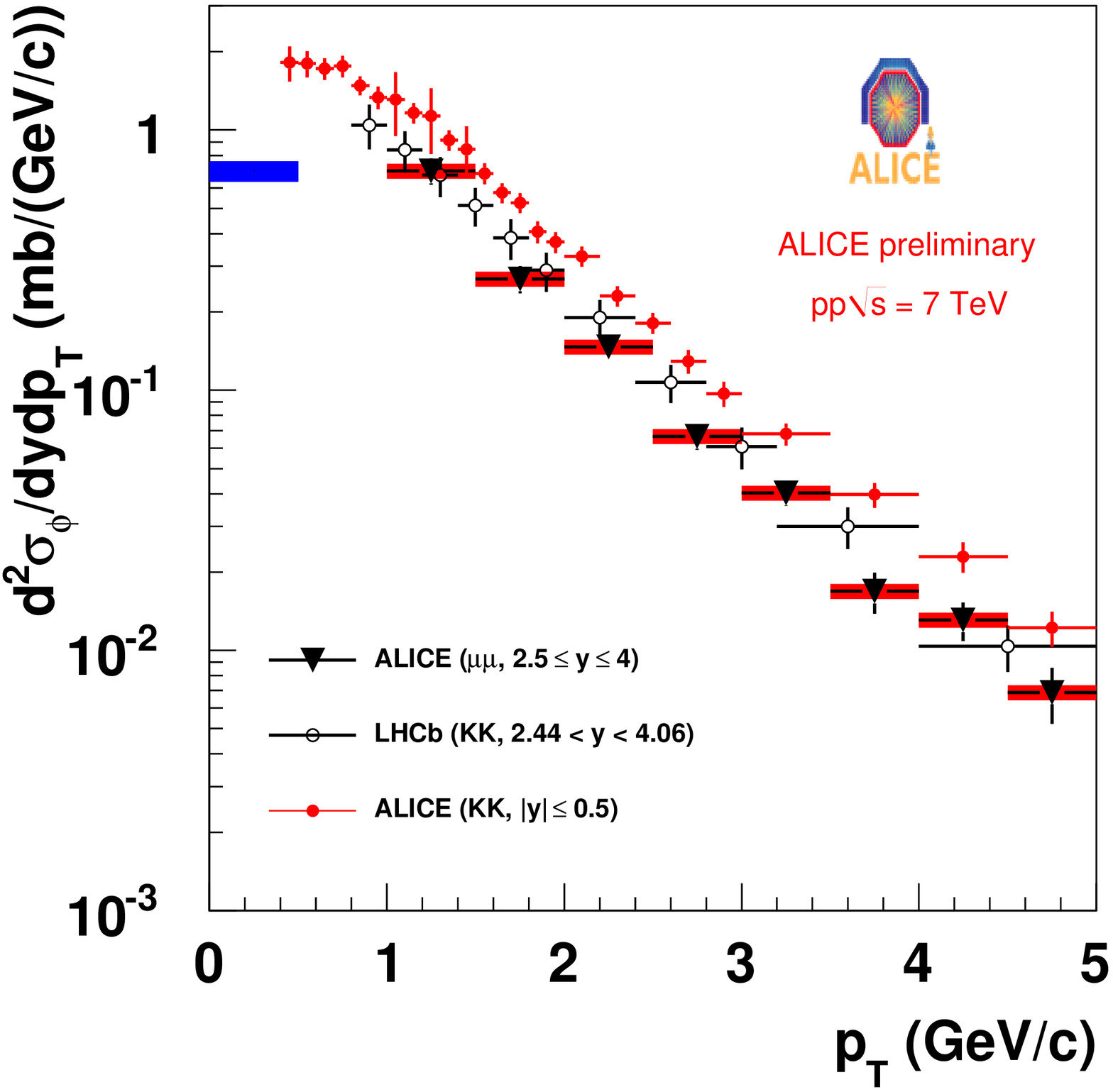}
	\caption{Left: dimuon invariant mass spectrum for $p_{\rm T}>1$~GeV$/c$.
	Yellow band: systematic uncertainty from background subtraction. 
	Red band: uncertainty in the relative normalization of the sources. 
		Right:
		${\rm d}^2\sigma_\phi/{\rm d}p_{\rm T} {\rm d}y$ in dimuons compared 
		to the LHCb~\cite{LHCb} and ALICE~\cite{ALICEKK} measurements in kaons.
	}
\label{fig:phi_all}
\end{figure}

After subtracting the combinatorial background from the opposite sign mass spectrum, 
we obtain the signal mass spectrum shown in Fig.~\ref{fig:phi_all} (left). 
%
The invariant mass spectrum is fitted with the contributions given by the 
light meson decays into muons and open charm/beauty contributions. 
The free parameters of the fit are the normalizations of 
the $\eta\rightarrow\mu \mu \gamma$, $\omega\rightarrow\mu \mu$, 
$\phi\rightarrow\mu \mu$ and open charm signals. The other
processes ($\eta \rightarrow \mu\mu$, 
$\rho \rightarrow \mu\mu$, 
$\omega \rightarrow \mu\mu \pi^0$, 
$\eta' \rightarrow \mu\mu \gamma$ and open beauty)
are fixed according to the relative branching ratios or 
cross sections.
The main sources of systematic uncertainty are due to the uncertainty in the 
background normalization and on the relative normalization of the sources, 
mainly due to the error on the branching ratios for
the $\omega$ and $\eta'$ Dalitz decays.
The raw number of $\phi$ and $\rho+\omega$ resonances obtained from the fit is
$N_\phi=(3.20 \pm 0.15) \times 10^3$ and $N_{\rho+\omega}=(6.83 \pm 0.15) \times 10^3$. 

%
%
The $\phi$ production cross section was evaluated in the range 
$2.5 < y < 4$, $1 < p_{\rm T} < 5$~GeV/$c$ through the formula
$\sigma_\phi={N_\phi^c \over {BR(\phi \rightarrow l^+ l^-)}} {\sigma_{MB} 
\over N_{MB}} {N^{MB}_\mu \over N^{\mu-MB}_\mu}$, 
where $N_\phi^c$ is the measured number of $\phi$ mesons 
corrected for the efficiency and the acceptance, 
$BR(\phi \rightarrow l^+ l^-) = (2.95 \pm 0.03) \times 10^{-4}$
is the branching ratio in lepton pairs, obtained as a weighted 
average of the branching ratios in $\rm e^+e^-$ and $\mu^+\mu^-$ 
pairs~\cite{PDG}, $N_{MB}$ is the number of minimum bias collisions,
$\sigma_{MB}$ is the ALICE minimum bias cross section in pp collisions
at $\sqrt{s}=7$~TeV and 
${N^{MB}_\mu / N^{\mu-MB}_\mu}$ is the ratio between the number 
of single muons in the region $2.5 < y_\mu < 4$, 
$p_{T,\mu}> 1$~GeV/$c$ collected with the minimum bias trigger and 
with the muon trigger. 
The minimum bias cross section was measured in a Van der Meer scan
~\cite{sigv0}. Its value is 
$\sigma_{MB}=62.3 \pm 0.4 (\mathrm{stat}) \pm 4.3 \mathrm{(syst)~mb}$.
The number of minimum bias collisions was corrected, run by run, for the 
probability of having multiple interactions in a single bunch crossing. 
The ratio ${N^{MB}_\mu / N^{\mu-MB}_\mu}$ strongly depends on the 
data taking conditions and was evaluated run by run. 
We obtain $\sigma_\phi(1<p_{\rm T}<5~{\rm GeV}/c,2.5<y<4)=0.940 \pm  0.084 (\mathrm{stat}) \pm 0.095 \mathrm{(syst)~mb}$.
The systematic error comes from the uncertainty on the background subtraction
($2 \%$), the muon trigger efficiency ($4\%$), the tracking efficiency ($3\%$)
the uncertainty on the $\phi$ branching ratio into dileptons ($1\%$), 
on the minimum bias cross section ($7\%$) 
and on the ratio ${N^{MB}_\mu / N^{\mu-MB}_\mu}$ ($3\%$).

The $p_{\rm T}$-differential cross section 
${\rm d}^2\sigma_\phi/{\rm d}p_{\rm T} {\rm d}y$ 
is shown in Fig.~\ref{fig:phi_all} (triangles). Point to
point uncorrelated systematic uncertainties are indicated as red boxes. 
The fully correlated systematic uncertainty, represented by a blue box on the left 
side of the plot, amounts to $9\%$. A fit to the differential cross section 
with a power law function, $C \cdot p_{\rm T} / [1 + (p_{\rm T}/p_0)^2]^n$, gives 
$p_0=1.16 \pm 0.23$~GeV/$c$ and $n=2.7\pm 0.2$.

The measurements in kaon pairs 
performed by LHCb in a similar rapidity range ($2.44 < y < 4.06$, open circles)~\cite{LHCb}
and by ALICE at midrapidity ($|y|<0.5$, full circles)~\cite{ALICEKK} are also plotted, showing 
that the shapes are similar. 
The rescaling of the LHCb cross section to $p_{\rm T}>1$ GeV$/c$ and to $2.5<y<4$ leads to  
$\sigma_\phi=1.07\pm 0.15(\mathrm{full~error})$~mb.
There is a $14\%$ difference between the ALICE and LHCb measurements. 
Considering the ALICE statistical error and the part of the systematic uncertainty
which are certainly not correlated among the two experiments,
the two measurements are in agreement. 

\begin{figure}
	\centering
	\includegraphics[width=0.49 \textwidth]{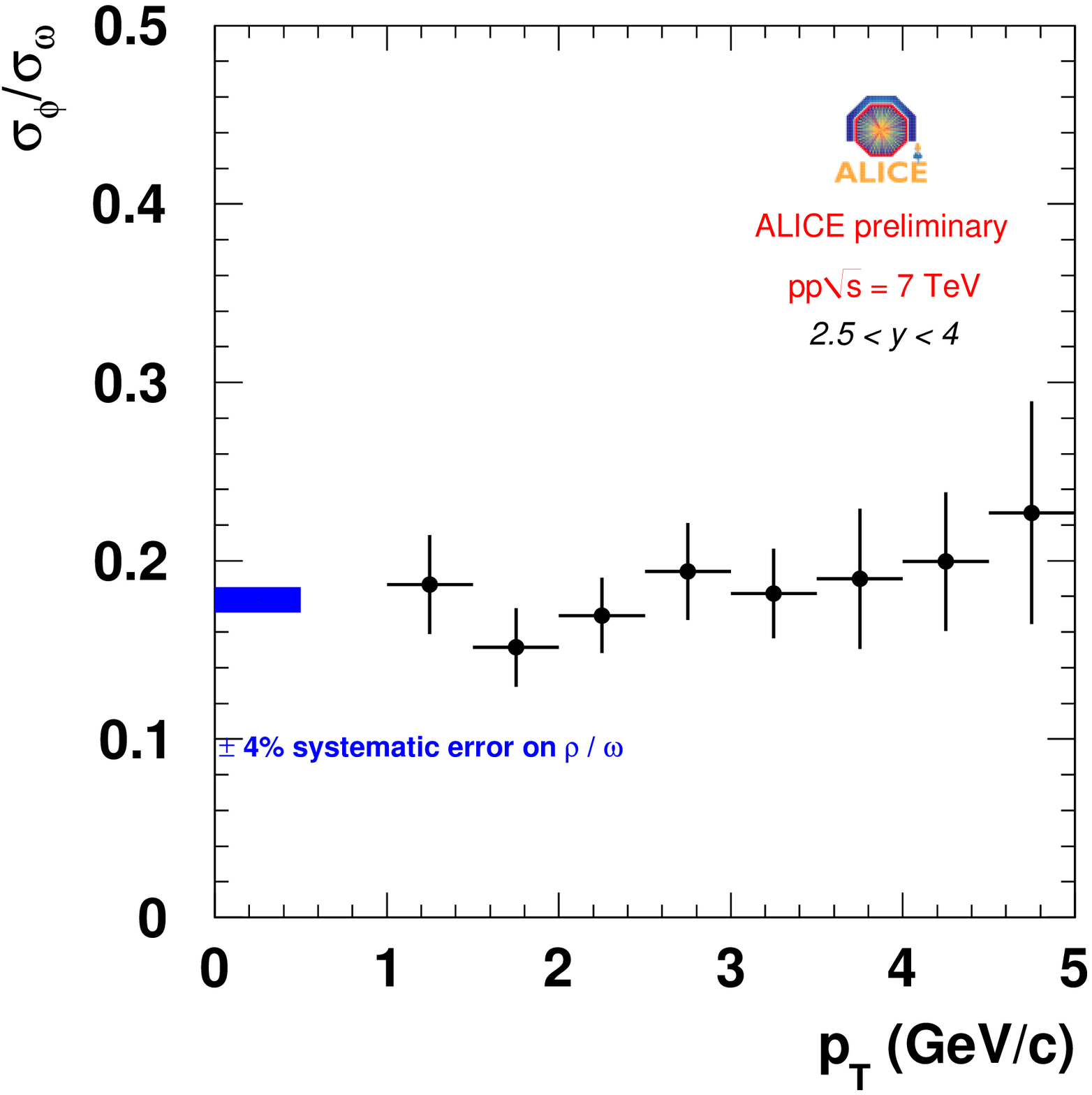}
	\includegraphics[width=0.49 \textwidth]{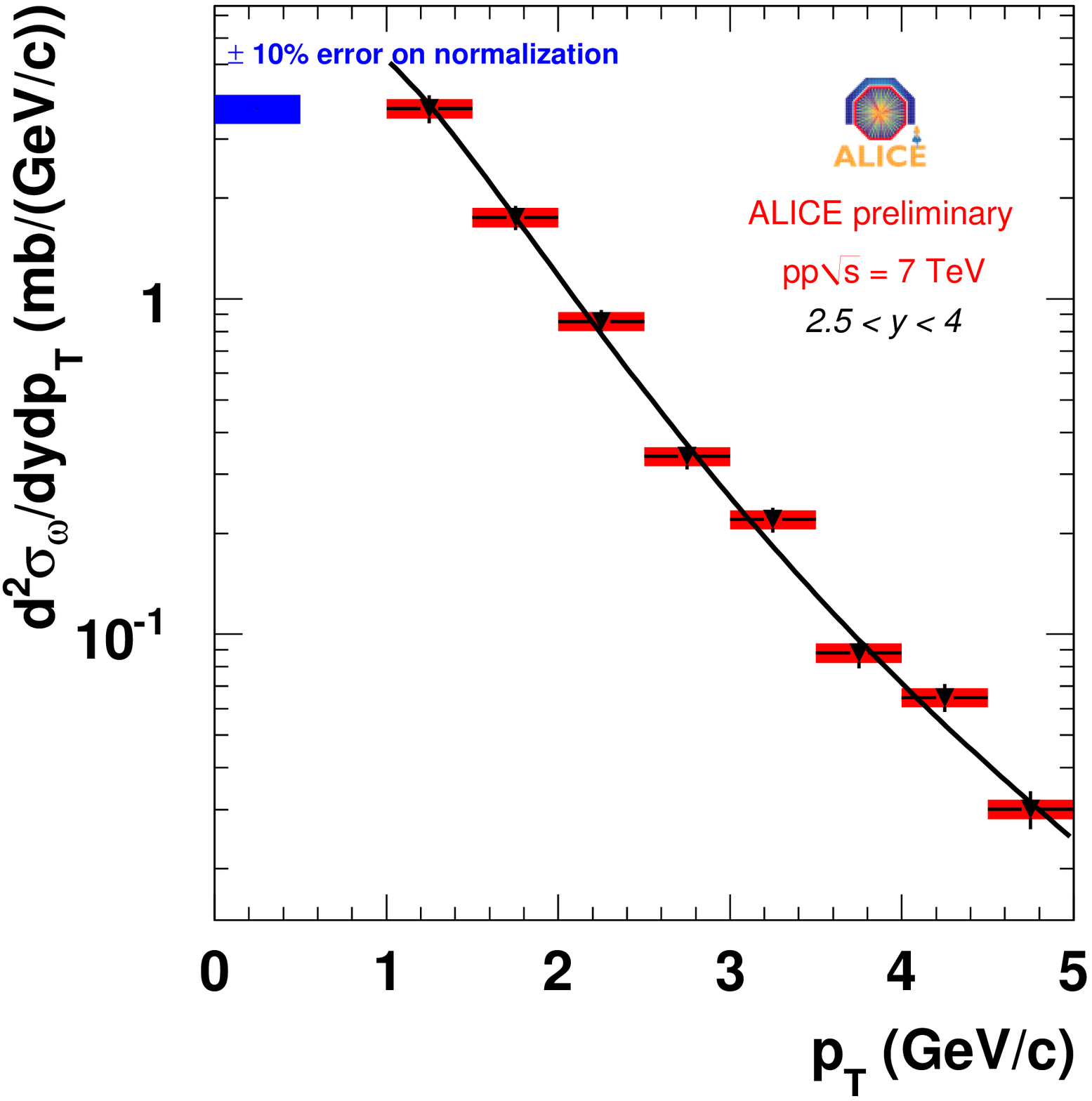}
	\caption{Left: Ratio $\sigma_\phi / \sigma_\omega$  
	as a function of $p_{\rm T}$. Right: 
	${\rm d}^2\sigma_\omega/{\rm d}p_{\rm T} {\rm d}y$ for $2.5 < y < 4$.}
\label{fig:dsigmaomegadydpt}
\end{figure}
The ratio 
$N_\phi/(N_\rho+N_\omega)$
was measured as a function of the transverse 
momentum, showing a flat trend with an average value of $0.42 \pm 0.02$. 
In order to extract the $\omega$ cross section, the $\rho$ and $\omega$ contributions 
must be disentangled, leaving the $\rho$ normalization as an additional free parameter 
in the fit to the dimuon mass spectrum. The result of the fit gives 
$\sigma_\rho / \sigma_\omega = 1.15 \pm 0.20 \mathrm{(stat)} \pm 0.12 \mathrm{(syst)}$.
The systematic uncertainty was evaluated changing the normalizations of the 
$\eta' \rightarrow \mu \mu \gamma$ and $\omega \rightarrow \mu \mu \pi^0$ 
according to the uncertainties in their branching ratios, and the 
background level by $\pm 10\%$, twice the uncertainty in the normalization. 
From these results, it was possible to extract the ratio 
$\sigma_\phi / \sigma_\omega = 0.178 \pm 0.015 \mathrm{(stat)} \pm 0.008 \mathrm{(syst)}$. 
This ratio is plotted as a function of $p_{\rm T}$ in Fig.~\ref{fig:dsigmaomegadydpt} (left).  
The $\omega$ production cross section, calculated from this ratio, is 
$\sigma_\omega (1<p_{\rm T}<5~{\rm GeV}/c,2.5<y<4) = 5.28 \pm 0.46 \mathrm{(stat)} \pm 0.58 \mathrm{(syst)~mb}$. 
In Fig.~\ref{fig:dsigmaomegadydpt} (right) the $\omega$ differential cross section is shown. 
Data are fitted with the power law function, obtaining
$p_0=1.44 \pm 0.09$~GeV/$c$ and $n=3.2\pm 0.1$.
%
%

In conclusion, the $\phi$ and $\omega$ $p_{\rm T}$ differential cross sections were measured in 
pp collisions at $\sqrt{s}=7$~TeV. The ratio between the $\phi$ and the $\omega$ cross 
sections is flat as a function of $p_{\rm T}$. 
In Pb-Pb collisions work is in progress to measure $\phi$ production as a function 
of centrality. 

\end{document}